\documentclass[aps,pre,twocolumn,showpacs,floatfix]{revtex4}
\usepackage{amsmath,bm,epsfig}

\begin{document}

\title{Ultimate-state scaling in a shell model for
homogeneous turbulent convection}
\author{Emily S.C. Ching$^{1,2}$ and T.C. Ko$^1$}
\affiliation{$^1$ Department of Physics,
The Chinese University of Hong Kong, Shatin, Hong Kong}
\affiliation{$^2$ 
Institute of Theoretical Physics,
The Chinese University of Hong Kong, Shatin, Hong Kong}

\date{\today}
 
\begin{abstract}
An interesting question in turbulent convection is how the heat transport depends
on the strength of thermal forcing in the limit of very large thermal forcing. 
Kraichnan predicted [Phys. Fluids {\bf 5}, 1374 (1962)] that the heat 
transport measured by the Nusselt number (Nu) would depend on the strength of thermal 
forcing measured by the Rayleigh number (Ra) as Nu $\sim$ Ra$^{1/2}$ with possible
logarithmic corrections at very high Ra. This scaling behavior is taken as a signature of the 
so-called ultimate state of turbulent convection. The ultimate state
was interpreted in the Grossmann-Lohse (GL) theory [J. Fluid Mech. {\bf 407}, 27 (2000)] 
as a bulk-dominated state in which both the kinetic and thermal dissipation are 
dominated by contributions from the bulk of the flow with the boundary layers either 
broken down or playing no role in the heat transport. In this paper,
we study the dependence of Nu and the Reynolds number (Re) measuring the 
root-mean-squared velocity fluctuations on Ra and the Prandtl number (Pr) 
using a shell model for homogeneous turbulent convection where buoyancy 
is acting directly on most of the scales.  We find that
Nu$\sim$ Ra$^{1/2}$Pr$^{1/2}$ and Re$\sim$ Ra$^{1/2}$Pr$^{-1/2}$, which
resemble the ultimate-state scaling behavior for fluids with moderate Pr, but the presence of a drag 
acting on the large scales is crucial in giving rise to such scaling. 
This suggests that if buoyancy acts on most of the scales in the bulk of turbulent 
convection at very high Ra, then the ultimate state cannot be a bulk-dominated state.
\end{abstract}
\maketitle

\section{Introduction}

In Rayleigh-B{\'e}nard convection, fluid confined in a box is heated from below 
and cooled on top. When the temperature difference is large enough, convective 
motion sets in. The flow state is characterized by the geometry of the box and 
two control parameters: the Rayleigh number (Ra), which measures the strength of
the thermal forcing and the Prandtl number (Pr), which is the
ratio of the diffusivities of momentum and heat of the fluid.
The two parameters are defined by
Ra $= \alpha g \Delta L^3/ (\nu \kappa)$, and
Pr $= \nu/\kappa$, where $\Delta$ is the temperature
difference, $L$ is the height of the box,
$g$ the acceleration due to gravity, and $\alpha$,
$\nu$, and $\kappa$ are respectively the volume expansion coefficient,
kinematic viscosity and thermal diffusivity of the fluid.
When Ra is sufficiently large, the convective motion becomes turbulent.
Turbulent Rayleigh-B{\'e}nard convection has been a system of great research 
interest~(see, for example,~\cite{Siggia,Kadanoff} for a review). In particular,
an interesting question is how the heat transport of the fluid, measured by the Nusselt
number (Nu), which is defined as the actual heat transport normalized by 
that when there were only pure conduction, depends on Ra in the limit of very high Ra. 
More than 40 years ago, Kraichnan predicted~\cite{Kraichnan} 
that in this asymptotic limit, 
\begin{eqnarray}
{\rm Nu} &\sim& {\rm Ra}^{1/2} (\ln {\rm Ra})^{-3/2} {\rm Pr}^{1/2} 
\label{Kraichnan1Nu} \\
{\rm Re}_0 &\sim& {\rm Ra}^{1/2} (\ln {\rm Ra})^{-1/2} {\rm Pr}^{-1/2}
\label{Kraichnan1Re}
\end{eqnarray}
for Pr $< 0.15$ and
\begin{eqnarray}
{\rm Nu} &\sim& {\rm Ra}^{1/2} (\ln {\rm Ra})^{-3/2} {\rm Pr}^{-1/4} 
\label{Kraichnan2Nu} \\
{\rm Re}_0 &\sim& {\rm Ra}^{1/2} (\ln {\rm Ra})^{-1/2} {\rm Pr}^{-3/4} 
\label{Kraichnan2Re}
\end{eqnarray}
for $0.15 <{\rm Pr} \le 1$.
Here Re$_0 = u_0 L/(2\nu)$ is the Reynolds number measuring the root-mean-squared 
horizontal velocity fluctuations $u_0$ at mid-height $L/2$.
Such a scaling behavior of Nu $\sim$ Ra$^{1/2}$ and Re$_0$~$\sim$Ra$^{1/2}$ 
is taken to be a signature
of the so-called ultimate state of turbulent convection.
This predicted asymptotic increase of Nu as Ra$^{1/2}$ is 
stronger than the observed dependence of Ra$^{\gamma}$ with $\gamma$ around 0.3 
at moderate Ra. According to Kraichnan, the 
convective eddies, produced in the bulk of the turbulent convective flow, 
generate turbulent shear boundary layers near the walls and it is the
small-scale turbulence present in these boundary layers that dominates and enhances 
the heat transport.
Thus, the shear boundary layers play a crucial role in giving rise to the ultimate-state
scaling in Kraichnan's work. 

On the other hand, a recent theory proposed by Grossmann and Lohse~\cite{GL}
agrued that the kinetic and thermal boundary layers would 
either break down or do not contribute to the energy and thermal dissipation and 
thus do not play any role in the heat transport at very high Ra. 
In this bulk-dominated state, the Grossmann-Lohse (GL) theory predicted that
for fluids with moderate Pr:
\begin{eqnarray}
{\rm Nu} &\sim& {\rm Ra}^{1/2} {\rm Pr}^{1/2}  
\label{GLbulkNu} \\
{\rm Re}_{LSC} &\sim& {\rm Ra}^{1/2} {\rm Pr}^{-1/2} 
\label{GLbulkRe}
\end{eqnarray}
where Re$_{LSC}= UL/\nu$ is the Reynolds number measuring the mean large-scale circulating 
flow velocity $U$ near
the boundaries. Thus GL predicted the same scaling of
Ra$^{1/2}$ for Nu and Re$_{LSC}$ for fluids with moderate Pr. 
The Pr-dependence predicted in the GL theory agrees with that of Kraichnan 
for Pr $<0.15$ but not for $0.15 < {\rm Pr} \le 1$. 
We emphasize that although the predicted asymptotic dependence of Nu and
Re$_{LSC}$ or Re$_0$ are both Ra$^{1/2}$ in the two theories,
the assumed roles of the boundary layers in heat 
transport in the asymptotic regime are rather different.

The ultimate state of turbulent convection has been elusive~\cite{NatureView} in that
definitive experimental evidence is lacking.
An increase in the Nu-Ra scaling exponent was found around Ra=$10^{11}$ 
by Chavanne {\it et al.}~\cite{Grenoble1,Grenoble2} in experiments using low 
temperature helium gas, and was
interpreted as the transition to the ultimate state.
However, similar experiments by Niemela {\it et
al.}~\cite{SreeniNature} showed that the measurements of Nu can be well 
represented by Nu~$\sim$~Ra$^{0.309}$ for Ra up to around $10^{17}$. 
This puzzling discrepancy in the two experiments
remains unresolved. The situation is further complicated by  
the increasing difficulty to keep the experiments 
within the Boussinesq approximation at high Ra~\cite{Sreeni}.

GL's work led to the idea of attaining the ultimate-state scaling at moderate Ra by
an artificial destruction of the boundary layers~\cite{LTPRL2003}.
By numerically simulating the bulk of 
turbulent Rayleigh-B{\'e}nard convective flow, modeled by 
three-dimensional homogeneous turbulent convection with periodic boundary conditions~\cite{Orszag}, 
Lohse and coauthors~\cite{LTPRL2003,bulkPF2005} 
reported results that are consistent with Eqs.~(\ref{GLbulkNu}) and (\ref{GLbulkRe}),
when Re$_{LSC}$ is replaced by the Reynolds number measuring 
the root-mean-squared velocity fluctuations.
It has been found that~\cite{Orszag,Biferale}. 
for three-dimensional homogeneous turbulent convection,
buoyancy is relevant only at the largest scales. 
On the other hand the Bolgiano length~\cite{B59}, which is an estimate
of the length scale above which buoyancy forces are dominant, increases with Ra at moderate
Ra. Thus it is unclear whether or not buoyancy remains to be relevant
only at the largest scales at very high Ra. 

It is therefore interesting to study the scaling of heat transport using a model 
for homogeneous turbulent convection in which buoyancy is acting directly on most of the scales.
In this paper, we perform such a study
using a shell model for homogeneous turbulent convection, 
and buoyancy is acting on most of the scales in some parameter range.
We investigate
the scaling behavior of Nu and the Reynolds number (Re) measuring 
the root-mean-squared velocity fluctuations. The paper is organized as follows.
In Sec.~\ref{shellmodel}, we describe the shell model for homogeneous turbulent convection 
used in the present study, define Nu, Re and Ra in the model, and derive two exact results.
We present and discuss our results of the 
dependence of Nu and Re on Ra and Pr in Sec.~\ref{results}. 
In Sec.~\ref{epsilon}, we understand why our observed result of 
the average energy dissipation $\epsilon$ is different from that predicted in GL.
With the relative simplicity of the shell model, we can derive analytically results
for the scaling behavior of Nu and Re. We will illustrate this in Sec.~\ref{estimates},
and also show that the theoretical results agree well with the numerical observations.
Finally, we summarize and conclude in Sec.~\ref{summary}.

\section{The shell model for homogeneous turbulent convection}
\label{shellmodel}

Homogeneous turbulent convection, which represents the bulk of turbulent Rayleigh-B{\'e}nard
convection, has been proposed~\cite{Orszag} as a three-dimensional convective flow in a box,
with periodic boundary conditions, driven by a constant temperature gradient along the
vertical direction. In Boussinesq approximation~\cite{Landau},
the equations of motion read~\cite{LTPRL2003}:
\begin{eqnarray}
\frac{\partial {\vec u}}{\partial t}
+ {\vec u} \cdot {\vec \nabla} {\vec u}
&=& -{\vec \nabla} p + \nu \nabla^2 {\vec u} + \alpha g \theta {\hat z}
\label{ueqn}  \\
\frac{\partial \theta}{\partial t}
+ {\vec u} \cdot {\vec \nabla} \theta &=& \kappa \nabla^2 \theta + \beta u_z
\label{Teqn}
\end{eqnarray}
with ${\vec \nabla} \cdot {\vec u} = 0$.
Here, ${\vec u}$ is the velocity, $p$ is
the pressure divided by the density,
$\theta = T- (T_0 - \beta z)$ is the deviation of temperature $T$ from
a linear profile of constant temperature gradient of $-\beta$,
$T_0$ is the mean temperature, and
${\hat z}$ is a unit vector in the vertical direction.
A dynamical shell model for this system
has been proposed by Brandenburg~\cite{Brandenburg}.
Shell model is constructed in a discretized Fourier space with
$k_n= k_0 h^n$, $n = 0, 1, \ldots, N-1$, being the wavenumber in the $n$th shell,
and $h$ and $k_0$ are customarily taken to be 2 and 1 respectively.
Shell models for homogeneous and isotropic turbulence have been studied extensively
and proved to be successful in reproducing the scaling properties observed in
experiments~\cite{Shell}. In Brandenburg's model,
the velocity and temperature variables $u_n$ and $\theta_n$ are real
and satisfy the evolution equations:
\begin{eqnarray}
\nonumber
\frac{d u_n}{d t} 
&=& ak_n(u_{n-1}^2-hu_nu_{n+1}) 
 + bk_n(u_nu_{n-1}-hu^2_{n+1}) \\ 
&& -\nu k_n^2 u_n + \alpha g \theta_n
\label{un} \\
\nonumber
\frac{d \theta_n}{d t} 
&=&  \tilde ak_n(u_{n-1}\theta_{n-1}-hu_n\theta_{n+1})  \\
&+& \tilde b k_n(u_n \theta_{n-1}-hu_{n+1}\theta_{n+1})  
 -\kappa k_n^2 \theta_n + \beta u_n \ \
\label{thetan}
\end{eqnarray}
where $a$, $b$, $\tilde a$, and $\tilde b$ are positive parameters.

\begin{figure}[t]
\centering
\includegraphics[width=.40\textwidth,angle=270]{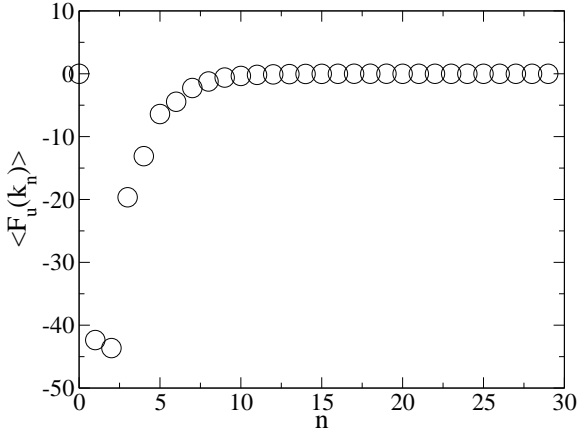}
\caption{$F_u(k_n)$ for $a=0.01$, $b=\tilde a=\tilde b=\alpha g=1$, $\beta=100$,
$\nu=\kappa=10^{-8}$.}
\label{fig1}
\end{figure}

For this shell model, it was found~\cite{ChingCheng} that
when $b/a$ is larger than some critical value around 2,
the effect of buoyancy is greater than the average energy dissipation rate for most shells.
Moreover, buoyancy directly affects the 
statistics of the system such that the scaling behavior of $u_n$ and $\theta_n$
is given by the Bolgiano-Obukhov scaling~\cite{B59,O59}~($u_n \sim k_n^{-3/5}$, $\theta_n \sim
k_n^{-1/5}$) plus corrections rather than Kolmogorov 1941 scaling~\cite{K41}~($u_n \sim
k_n^{-1/3}$, $\theta_n \sim k_n^{-1/3}$) plus corrections.
In other words, buoyancy is directly acting on most of the scales when $b/a$ is sufficiently large.
Thus we shall focus on $b/a$ large in the present work.
It was reported in earlier studies~\cite{Brandenburg} 
that the value of $b/a$ controls the direction of energy transfer. For
large $b/a$, there is an inverse energy transfer from
small to large scales.
The direction of energy transfer can be quantified by the sign of the 
energy transfer rate. Multiply Eq.~(\ref{un}) by $u_n$, we get
\begin{equation}
\frac{d E_n}{dt} = F_u(k_n)-F_u(k_{n+1})-\nu k_n^2 u_n^2 + \alpha g u_n \theta_n
\label{En}
\end{equation}
where $E_n = u_n^2/2$ is the energy in the $n$th shell and
\begin{equation}
F_u(k_n) \equiv k_n(au_{n-1}+bu_n)u_{n-1}u_n
\label{Fu}
\end{equation}
is the rate of energy transfer or energy flux from $(n-1)$th to $n$th shell.
As shown in Fig.~\ref{fig1},
$\langle F_u(k_n) \rangle$ is indeed negative, confirming that, on average, 
energy is transferred from large $n$ (small scales) to small $n$ (large scales)
when $b/a$ is large. As a result,
a drag acting on the largest scale has to be added to dampen the
growth of energy at large scales so that the system can achieve a statistically
stationary state~\cite{2dPRE,CelaniPF,SuzukiPRE}.
Thus we modify Eq.~(\ref{un}) to
\begin{eqnarray}
\nonumber
\frac{d u_n}{d t} &=& ak_n(u_{n-1}^2-hu_nu_{n+1})
+ bk_n(u_nu_{n-1}-hu^2_{n+1}) \\
&& - \nu k_n^2 u_n + \alpha g \theta_n - f u_0 \delta_{n,0}
\label{undamp}
\end{eqnarray}
where $f u_0 \delta_{n,0}$, with $f>0$, 
is a linear drag term acting only on the first shell $n=0$.
Also, Eq.~(\ref{En}) becomes
\begin{equation}
\frac{d E_n}{dt} = F_u(k_n)-F_u(k_{n+1})-\nu k_n^2 u_n^2 + \alpha g u_n \theta_n - f u_0^2
\delta_{n,0}
\label{Endamp}
\end{equation}

Next we need to define Ra, Nu and Re in the shell model noting that
Pr is given by the usual definition of $\nu/\kappa$.
The definitions of Ra and Re are straightforward: we only need to replace $L$ by $1/k_0$ 
such that 
\begin{eqnarray}
{\rm Ra} &=& \frac{\alpha g \beta}{ k_0^4 \nu \kappa}
\label{Ra} \\
{\rm Re} &=& \frac{\left[\sum_n \langle u_n^2 \rangle\right]^{1/2}}{\nu k_0}
\label{Re}
\end{eqnarray}
As for Nu, we recall its definition in
turbulent Rayleigh-B{\'e}nard convection as:
\begin{equation}
{\rm Nu} \equiv \frac{\langle u_z (T-T_0) - \kappa \partial T/\partial z \rangle_A}{\kappa \Delta/L} 
= \frac{\langle u_z \theta \rangle_V}{\kappa \Delta/L}+1 
\end{equation}
where $\langle \cdots \rangle_A$ 
denotes an average over (any) horizontal plane of the convection cell and 
time, and 
$\langle \cdots \rangle_V$ denotes an average over the whole volume of the convection cell and time.
Thus we define
\begin{equation}
{\rm Nu} = \frac{\sum_n \langle u_n \theta_n \rangle}{\kappa \beta} + 1
\label{Nu}
\end{equation}
accordingly with $\langle \cdots \rangle$ denotes an average over time. 

One can derive two exact results in exact analogy to those derived 
in the case of turbulent Rayleigh-B{\'e}nard convection~\cite{Siggia}. 
Multiply Eq.~(\ref{thetan}) by $\theta_n$, we get
\begin{equation}
\frac{d S_n}{dt} = F_\theta(k_n)-F_\theta(k_{n+1})-\kappa k_n^2 \theta_n^2 +
\beta u_n \theta_n
\label{Sn}
\end{equation}
where
$S_n \equiv \theta_n^2/2$ is the entropy, which 
is proportional to the volume
integral of temperature fluctuations in Bousinessq approximation, in the $n$th shell and 
\begin{equation}
F_\theta(k_n) \equiv k_n({\tilde a}u_{n-1}+
{\tilde b}u_n)\theta_{n-1}\theta_n
\label{Ftheta}
\end{equation}
is the rate of entropy transfer or entropy flux from $(n-1)$th to $n$th shell.
In the statistically stationary state,
summing Eq.~(\ref{Endamp}) and Eq.~(\ref{Sn}) over $n$ and using
Eqs.~(\ref{Nu}) and (\ref{Ra}) give the exact results as
\begin{eqnarray}
\epsilon_{total} &=& \nu^3 k_0^4 \ ({\rm Nu}-1) \ {\rm Ra} \ {\rm Pr}^{-2} 
\label{exactepsilon}\\
\chi &=& \kappa \beta^2 \ {\rm Nu}
\label{exactchi}
\end{eqnarray}
where the total energy dissipation rate $\epsilon_{total}$ is given by
\begin{equation}
\epsilon_{total} = \epsilon + \epsilon_{drag} 
\end{equation}
Here $\epsilon$ is the average energy dissipation rate given by
\begin{equation}
\epsilon \equiv \nu \sum_n k_n^2 \langle u_n^2 \rangle
\end{equation} 
and $\epsilon_{drag}$ is
the average rate of energy dissipation due to the large-scale drag:
\begin{equation}
\epsilon_{drag} = f \langle u_0^2 \rangle
\label{drag}
\end{equation}
The average thermal dissipation
rate $\chi$ is defined as
\begin{equation}
\chi \equiv \kappa \sum_n k_n^2 \langle \theta_n^2 \rangle + \kappa \beta^2
\end{equation}
in accordance with
$\chi \equiv \kappa \langle (\nabla T)^2 \rangle_V
= \kappa \langle (\nabla \theta)^2 \rangle_V + \kappa \beta^2$ for 
turbulent Rayleigh-B{\'e}nard convection.

\section{Results for ${\rm Nu(Ra,Pr)}$ and ${\rm Re(Ra,Pr)}$}
\label{results}

We numerically integrate Eqs.~(\ref{thetan}) and (\ref{undamp}) using 
fourth-order Runge-Kutta method with an initial condition of $u_n=\theta_n=0$ except for a small
perturbation of $\theta_n$ in an intermediate value of $n$. 
We use $a=0.01$, $b=\tilde a=\tilde b=\alpha g=1$,
$\beta = 100$, $f=0.5$, $N=30$, and vary $\nu$ and $\kappa$ to study Nu and Re as a 
function of Ra and Pr. Five moderate values of Pr ranging from 
0.1 to 2 are studied.

\begin{figure}[bth]
\centering
\includegraphics[width=.42\textwidth]{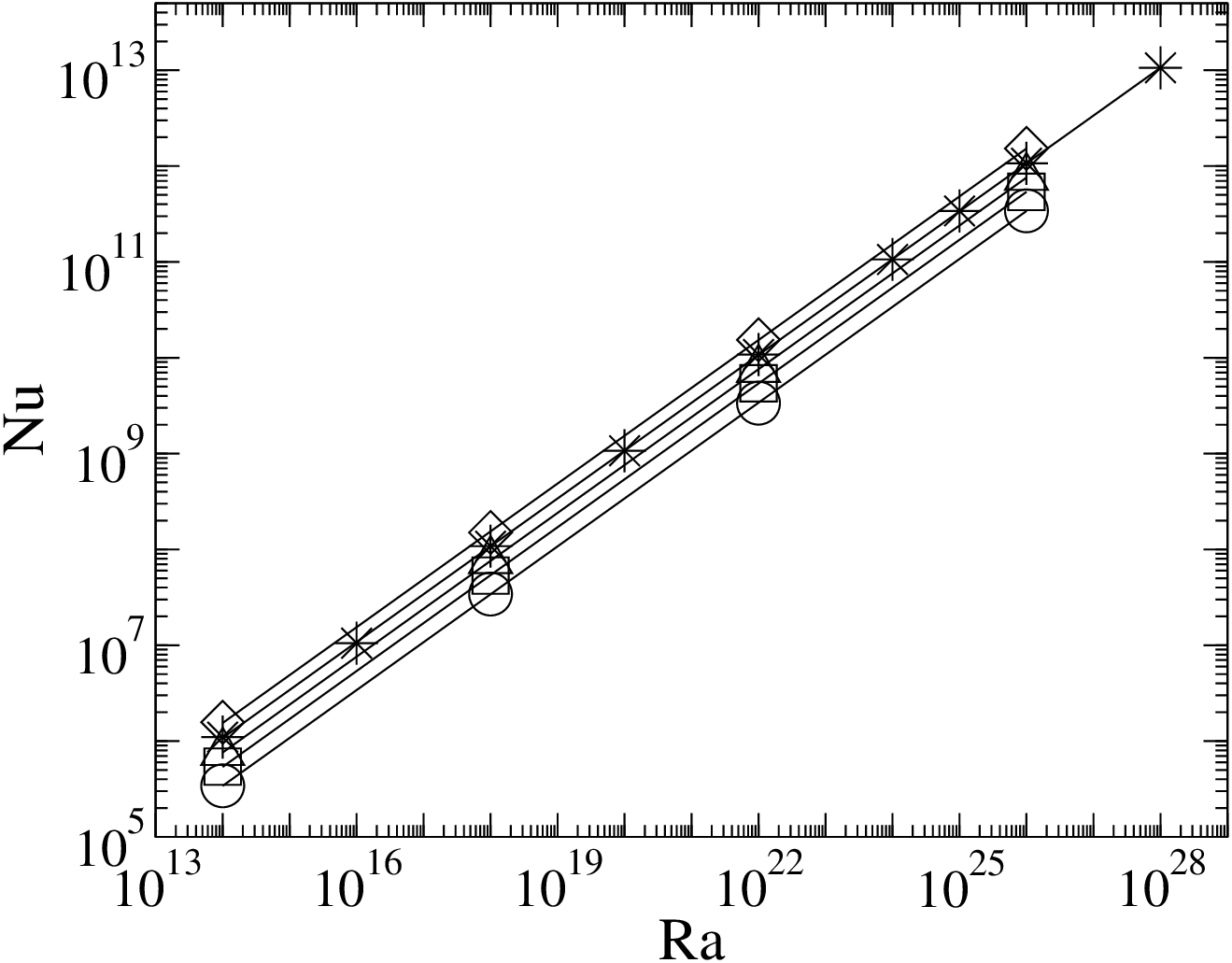}
\caption{Dependence of Nu on Ra for
for Pr = 0.1 (circles), Pr=0.25~(squares),
Pr=0.5~(triangles), Pr=1~(stars), and Pr=2~(diamonds).
The solid lines are the linear least-square fits of log Nu versus log Ra.}
\label{fig2}

\vspace{0.95cm}

\includegraphics[width=.4\textwidth]{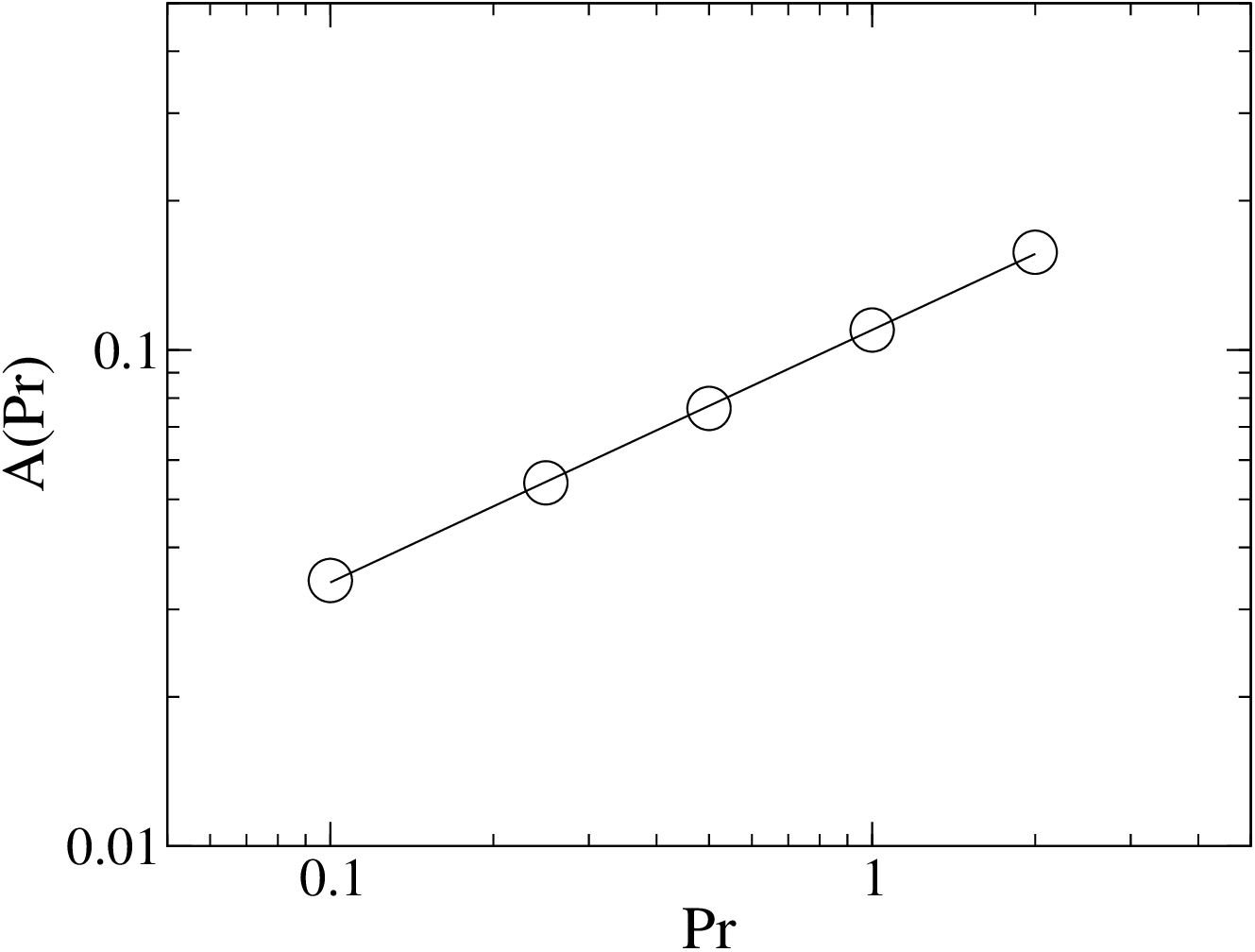}
\caption{The prefactor $A({\rm Pr})$ as a function of Pr. The solid line is the
linear least-square fit of log~$A$ versus log~Pr.}
\label{fig3}
\end{figure}

The dependence of Nu on Ra for each Pr are shown in Fig.~\ref{fig2}.
It can be seen that for each Pr, Nu scales with Ra:
${\rm Nu} = A({\rm Pr}){\rm Ra}^{\gamma_1}$
with $\gamma_1 = 0.500 \pm 0.001$. The Pr-dependence of Nu is found to be:
$A({\rm Pr}) = C_1 {\rm Pr}^{\gamma_2}$
with $\gamma_2 = 0.51 \pm 0.01$, as shown in Fig.~\ref{fig3}.
Thus we have
\begin{equation}
{\rm Nu} = C_1 {\rm Ra}^{0.500 \pm 0.001} {\rm Pr}^{0.51 \pm 0.01}
\label{Nuresult}
\end{equation}

\vspace{0.25cm}

\begin{figure}[bth]
\centering
\includegraphics[width=.42\textwidth]{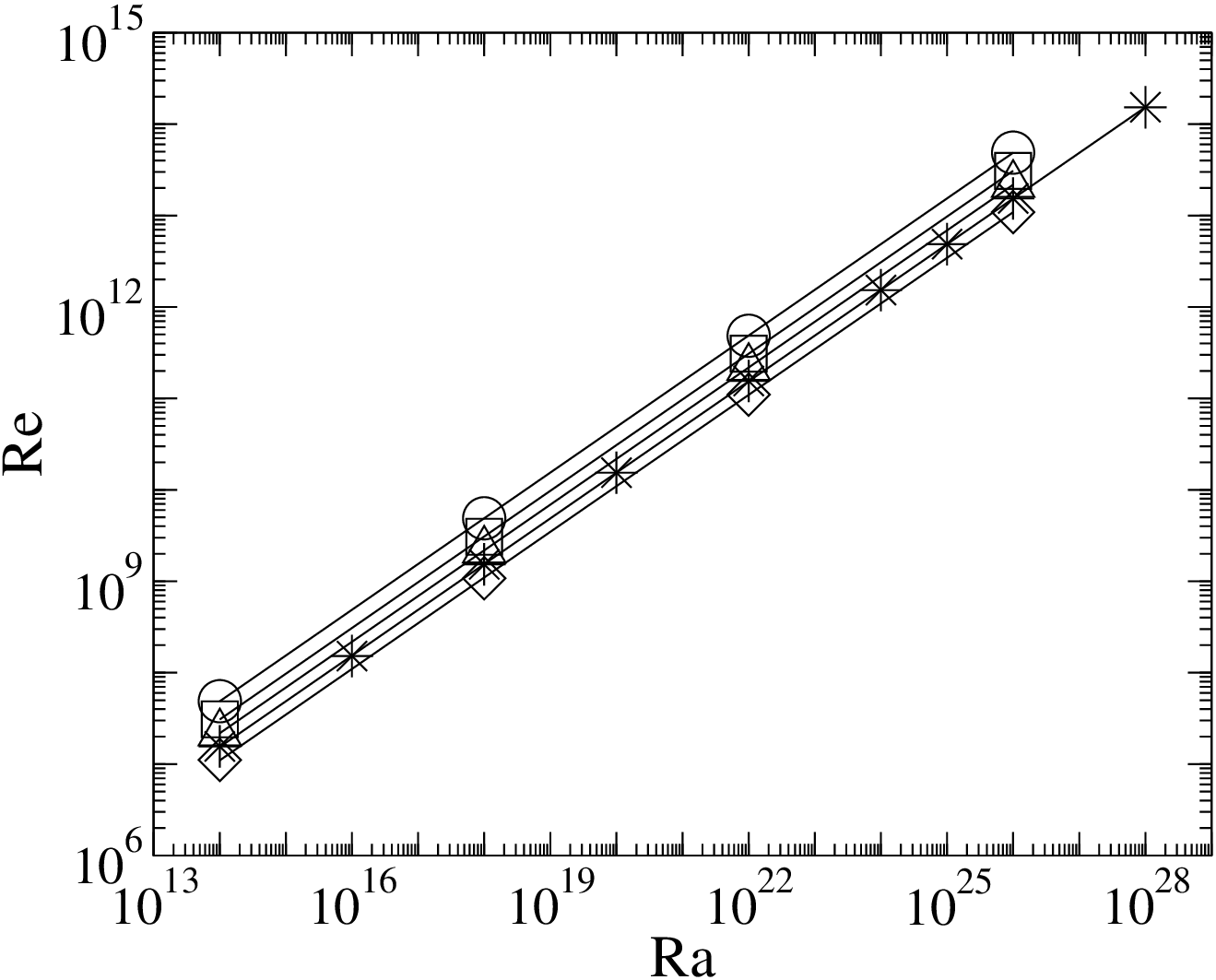}
\caption{Dependence of Re on Ra for different values of Pr,
with the same symbols as in Fig.~\ref{fig2}.
The solid lines are the linear least-square fits of log~Re versus log~Ra.}
\label{fig4}

\vspace{1.0cm}

\includegraphics[width=.42\textwidth]{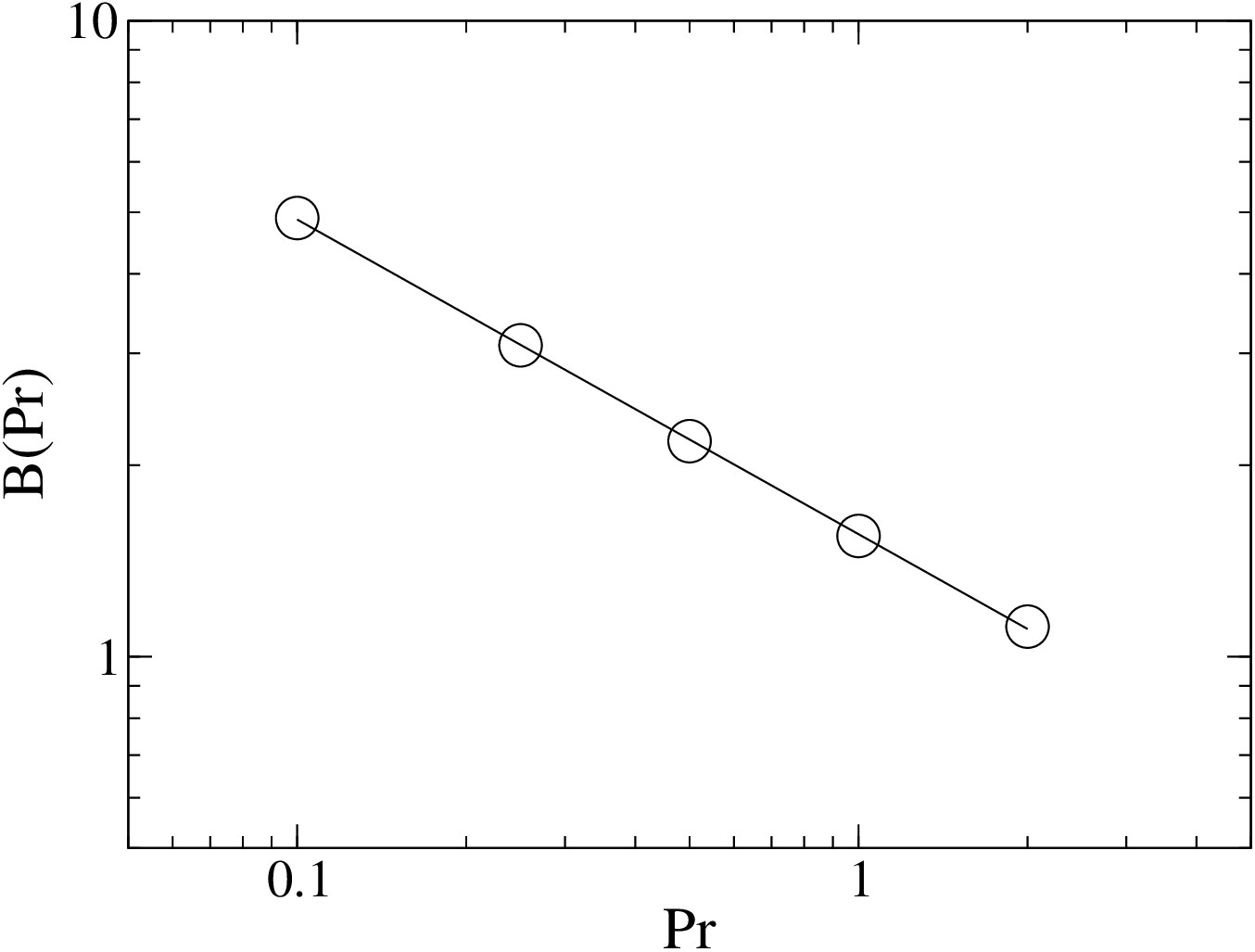}
\caption{The prefactor $B({\rm Pr})$ as a function of Pr. The solid line is the
linear least-square fit of log~$B$ versus log~Pr.}
\label{fig5}
\end{figure}

Similarly, we study the dependence of Re on Ra for each Pr.
As seen in Figs.~\ref{fig4} and
\ref{fig5}, Re = $B({\rm Pr})$Ra$^{\gamma_3}$ and $B({\rm Pr}) = C_2$ Pr$^{\gamma_4}$,
where $\gamma_3 = 0.500 \pm 0.001$ and $\gamma_4 = -0.50 \pm 0.01$. Thus
\begin{equation}
{\rm Re} = C_2 {\rm Ra}^{0.500 \pm 0.001} {\rm Pr}^{-0.50 \pm 0.01}
\label{Reresult}
\end{equation}
Hence our results for Nu and Re
are consistent with
Eqs.~(\ref{GLbulkNu}) and (\ref{GLbulkRe}), when
Re$_{LSC}$ is replaced by Re, and are also consistent with the numerical
results~\cite{LTPRL2003,bulkPF2005}
obtained in the three-dimensional homogeneous turbulent thermal convection
in which buoyancy only acts on the largest scales.

\vspace{0.5cm}

Moreover, we note that the Pr-dependence of
Nu and Re is consistent with GL and not with Kraichnan's result
for $0.15 < {\rm Pr} \le 1$ at very high Ra.
The significance of Nu~$\sim$~(Ra Pr)$^{1/2}$ and Re~$\sim$~(Ra/Pr)$^{1/2}$
is that the heat transport and the root-mean-squared velocity
fluctuations are independent of $\nu$ and $\kappa$. Thus for fluids with moderate Pr,
the scaling Nu~$\sim$~(RaPr)$^{1/2}$ at very high Ra is in line
of a `central dogma in turbulence'~\cite{NatureView}
that the effects of turbulence become independent of viscosity and 
thermal diffusivity when Re is sufficiently large. On the other hand,
the rigorous upper bound of Nu$\le 0.167$~Ra$^{1/2}-1$~\cite{DoeringConstantin},
for convection in a layer of fluid with no sidewalls, indicates~\cite{thank} that
the dependence of  Nu~$\sim$~(Ra Pr)$^{1/2}$ cannot hold for fluids with 
very large Pr.

\begin{figure}[hbt]
\centering
\includegraphics[width=.42\textwidth]{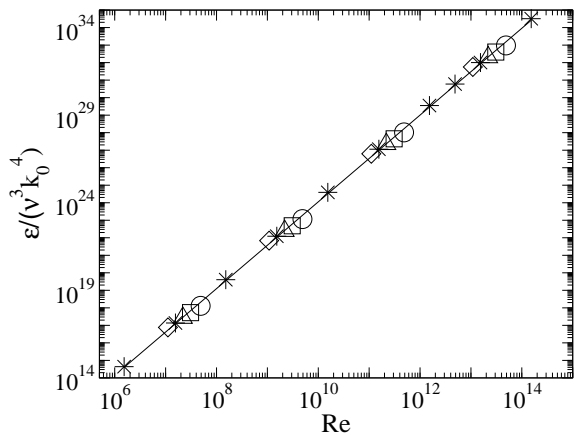}
\caption{$\epsilon/(\nu^3 k_0^4)$ as a function of Re for different values of
Pr with the same symbols as in Fig.~\ref{fig2}. 
The solid line is the linear least-square fit to all the data points 
in the log-log plot.}
\label{fig6}
\end{figure}

\begin{figure}[t]
\centering
\includegraphics[width=.45\textwidth]{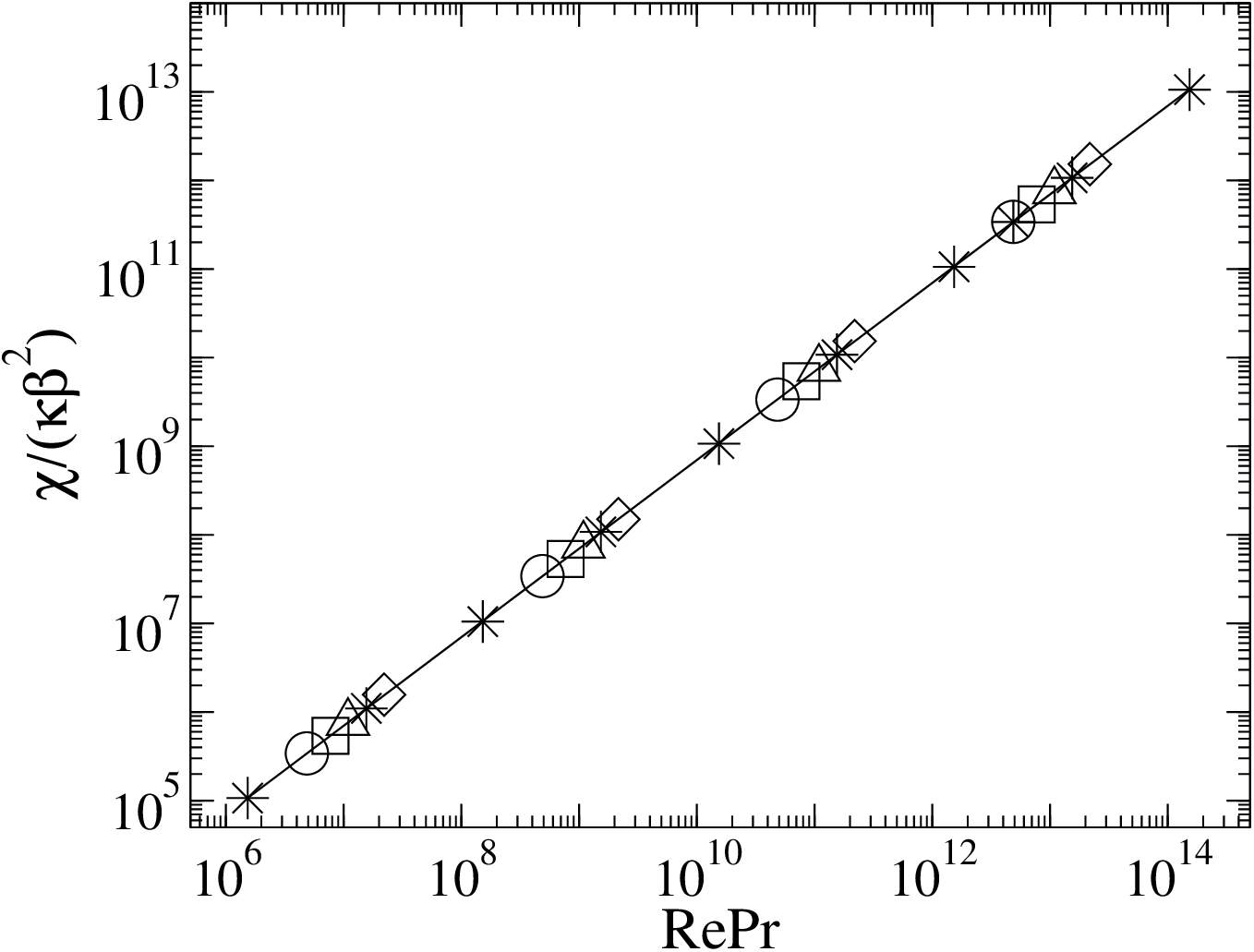}
\caption{$\chi/(\kappa \beta^2)$ as a function of RePr with the same
symbols as in Fig.~\ref{fig2}. The solid line is the linear least-square
fit to all the data points in the log-log plot.}
\label{fig7}

\vspace{1.5cm}

\includegraphics[width=.45\textwidth]{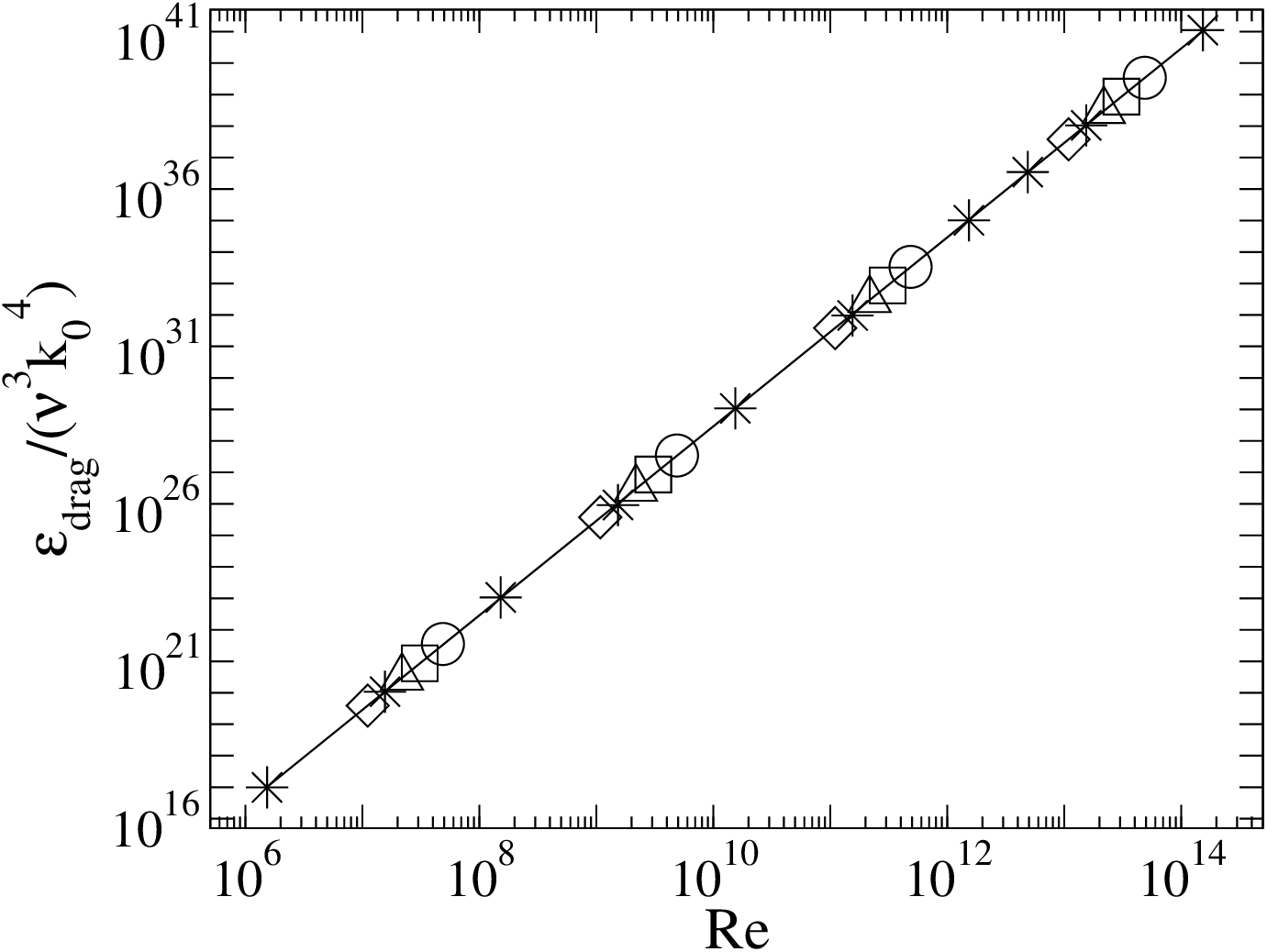}
\caption{$\epsilon_{drag}/(\nu^3 k_0^4)$ as a function of Re 
with the same symbols as in Fig.~\ref{fig2}. The solid line is the 
linear least-square fit 
to all the datapoints in the log-log plot.}
\label{fig8}
\end{figure}

In the derivation of Eqs.~(\ref{GLbulkNu}) and (\ref{GLbulkRe})
in the GL theory, there are two key intermediate
results, which
are the estimates of $\epsilon$ and $\chi$ in the bulk-dominated regime for moderate Pr.
In the shell model, $L \sim 1/k_0$, Re$_{LSC}$ becomes Re, and
these results translate to
\begin{eqnarray}
\epsilon^{(GL)} &\sim& \nu^3 k_0^4 \ {\rm Re}^3
\label{GLepsilonShell} \\
\chi^{(GL)} &\sim& \kappa \beta^2 \ {\rm Re Pr}
\label{GLchiShell}
\end{eqnarray}
Next we investigate the validity of
Eqs.~(\ref{GLepsilonShell}) and (\ref{GLchiShell}).
In Fig.~\ref{fig6}, we see that
\begin{equation}
\frac{\epsilon}{\nu^3 k_0^4} = C_3 {\rm Re}^{2.48 \pm 0.02}
\label{epsilonshell}
\end{equation}
Thus the dependence on Re is approximately Re$^{5/2}$ instead of Re$^3$.
On the other hand, as can be seen in Fig.~\ref{fig7}:
\begin{equation}
\frac{\chi}{\kappa \beta^2} = C_4 ({\rm Re Pr})^{1.000 \pm 0.001}
\label{chishell}
\end{equation}
in good agreement with Eq.~(\ref{GLchiShell}).

\vspace{1.cm}

\vspace{0.5cm}

The observed results of Eqs.~(\ref{epsilonshell}) and (\ref{chishell}) together 
with the exact results
Eqs.~(\ref{exactepsilon}) and (\ref{exactchi}) imply that $\epsilon_{total}$ cannot
be dominated by $\epsilon$ otherwise we would have
Re $\sim$ (Ra/Pr)$^{1/(\delta -1)}$ $\approx$ (Ra/Pr)$^{2/3}$,
which is in contradiction to our observed dependence of Re $\sim$ (Ra/Pr)$^{1/2}$.
Thus, $\epsilon_{total}$ has to be dominated by $\epsilon_{drag}$. Moreover, $\epsilon_{drag}$ 
has to scale as Re$^3$. Indeed we find that $\epsilon \ll \epsilon_{total}$ such that
\begin{equation}
\epsilon_{total} \approx \epsilon_{drag}
\label{epsilonTapprox}
\end{equation}
and as shown in Fig.~\ref{fig8},
\begin{equation}
\frac{\epsilon_{drag}}{\nu^3 k_0^4} = C_5 {\rm Re}^{3.00 \pm 0.01}
\label{epsilondrag}
\end{equation}
giving consistent results as expected.

Thus the observed scaling results of Nu and Re [Eqs.~(\ref{Nuresult}) and
(\ref{Reresult})]
depend crucially on the presence of a large-scale drag.
Such a damping mechanism at the largest scales cannot exist by itself in the bulk of
turbulentthermal convection but could be resulted from the interaction of the
boundaries with
the buoyancy-generated inverse transfer of energy from small to large scales.
This suggests that when buoyancy is acting directly on most scales,
the ultimate state, if exists, cannot
simply be a bulk-dominated flow state. Instead the boundaries must play a crucial
role.

\section{Understanding the dependence of $\epsilon$ on ${\rm Re}$} 
\label{epsilon}

In this section, we discuss how one can understand the observed dependence of 
$\epsilon$ on Re approximately as Re$^{5/2}$. 
The average energy dissipation rate in each shell increases with $n$ up to a maximum at the
dissipative scale, whose shell is denoted by the shell number $n_d$, then decreases again. Thus
$\epsilon$ can be approximated as :
\begin{equation}
\epsilon \approx D_1 \nu k_{n_d}^2 \langle u_{n_d}^2 \rangle
\label{epsilonapprox}
\end{equation}
where $D_1$ is a number of order 1. 
The dissipative wavenumber $k_{n_d}$ can be estimated as usual as
\begin{equation}
\frac{1}{k_{n_d}} = D_2 \left(\frac{\nu^3}{\epsilon}\right)^{1/4}
\label{kd}
\end{equation}
with $D_2$ being a number of order 1.
Now $\langle u_n^2 \rangle$ has good scaling behavior in $k_n$~\cite{Brandenburg,ChingCheng},
 for $0 \le n \le n_d$:
\begin{equation}
\langle u_n^2 \rangle \approx \langle u_0^2 \rangle \left(\frac{k_n}{k_0}\right)^{-2 \eta}
\label{unscaling}
\end{equation}
Moreover,
\begin{equation}
\sum_n \langle u_n^2 \rangle = D_3 \langle u_0^2 \rangle
\label{Reapprox}
\end{equation}
where $D_3$ is a number of order 1.
Putting Eqs.~(\ref{epsilonapprox})-(\ref{Reapprox}) together, we get
\begin{equation}
\frac{\epsilon}{\nu^3 k_0^4} \sim {\rm Re}^{\frac{4}{1+\eta}}
\label{epsilontheory}
\end{equation}

As discussed, when buoyancy is acting on most of the scales,
the scaling behavior is given by Bolgiano-Obukhov plus corrrections,
thus $\eta \approx 3/5$. This gives $4/(1+\eta) \approx 5/2$, as 
observed~[see Eq.~(\ref{epsilonshell})]. 
On the other hand when 
buoyancy is only acting on the largest scales, $u_n$ obeys
Kolmogorov 1941 scaling plus corrections~\cite{Brandenburg}. In this case,
$\eta \approx 1/3$, which leads to $4/(1+\eta) \approx 3$, giving the same 
intermediate result used in the GL theory~[see Eq.~(\ref{GLepsilonShell})].
Such a dependence of Re$^3$ can be easily understood
as follows. When buoyancy is only acting as a driving force at the largest scales, 
we have the usual energy cascade. From Eq.~(\ref{En}), we have
\begin{equation}
\epsilon \approx \nu k_{n_d}^2 \langle u_{n_d}^2 \rangle \approx \langle F_u(k_{n_d}) \rangle =
\langle F_u(k_1) \rangle \approx k_0 \langle u_0^2 \rangle^{3/2}
\label{epsilonestimate}
\end{equation}
using Eq.~(\ref{Fu}). Using Eq.~(\ref{Reapprox}), Eq.~(\ref{epsilonestimate}) 
leads immediately to $\epsilon \sim \nu^3 k_0^4$ Re$^3$.
When buoyancy is acting on most of the scales, 
there is, however, no longer a cascade of energy but only a cascade of entropy~\cite{ChingCheng}. 
In particular $\langle F_u(k_{n_d}) \rangle \ne \langle F_u(k_1) \rangle$ and, 
as a result, $\epsilon$ does not scale as Re$^3$.
In other words, the result Eq.~(\ref{GLepsilonShell}) estimated in the GL theory
does not hold when buoyancy is directly acting on most of the scales with the dynamics
governed by an entropy cascade.

\section{Estimates of $\chi$ and $\epsilon_{drag}$ and thus ${\rm Nu(Ra,Pr)}$ and 
${\rm Re(Ra,Pr)}$}
\label{estimates}

In the shell model,
there is always an entropy cascade
with $\langle F_\theta(k_n) \rangle \approx \chi$ 
being independent of $k_n$ in the intermediate scales. Thus
\begin{equation}
\chi \approx \langle F_\theta(k_1) \rangle \approx k_0 \langle u_0^2 \rangle^{1/2} \langle
\theta_0^2 \rangle  
\label{chiestimate}
\end{equation}
using Eq.~(\ref{Ftheta}).
On the other hand, Eq.~(\ref{Sn}) implies that
\begin{equation}
F_\theta(k_1) \approx \beta \langle u_0 \theta_0 \rangle \approx \beta \langle u_0^2 \rangle^{1/2}
\langle \theta_0^2 \rangle^{1/2}
\label{Fthetaestimate}
\end{equation}
Comparing Eqs.~(\ref{chiestimate}) and (\ref{Fthetaestimate}), we get
\begin{equation}
k_0 \langle \theta_0^2 \rangle^{1/2} \approx \beta
\label{beta}
\end{equation}
Putting Eq.~(\ref{beta}) into Eq.~(\ref{chiestimate}) and using Eq.~(\ref{Reapprox}), we obtain
\begin{equation}
\chi \approx \frac{\beta^2}{k_0} \sum_n \langle u_n^2 \rangle^{1/2}  
\label{chiE}
\end{equation}
which leads immediately to 
\begin{equation}
\chi \approx \kappa \beta^2 {\rm Re} \ {\rm Pr} 
\label{chiapprox}
\end{equation}
as observed~[see Eq.~(\ref{chishell})].

From Eqs.~(\ref{drag}) and (\ref{Reapprox}), we have
\begin{equation}
\epsilon_{drag} \sim f k_0^2 \nu^2 {\rm Re}^2 = \nu^3 k_0^4 \frac{f}{\sqrt{\alpha g \beta}} \ {\rm
Ra}^{1/2} {\rm Pr}^{-1/2} {\rm Re}^2
\label{dragapprox}
\end{equation}
For $\epsilon_{drag}/(\nu^3 k_0^4)$ to depend only on Ra and Pr, we require
\begin{equation}
f = h({\rm Ra},{\rm Pr}) \sqrt{\alpha g \beta}
\label{h}
\end{equation}
for some function $h$ of Ra and Pr when Ra and Pr are varied.
In our numerical calculations, $\alpha g$, $\beta$ and $f$ are kept fixed while $\nu$ and $\kappa$
are varied to get different Ra and Pr. This amounts to taking 
\begin{equation}
h({\rm Ra},{\rm Pr}) = h_0
\label{h0}
\end{equation}
for some fixed constant $h_0$.
Using Eqs.~(\ref{exactepsilon}), (\ref{epsilonTapprox}), (\ref{dragapprox}), (\ref{h}), and
(\ref{h0}), we get
\begin{equation}
{\rm Nu} \ {\rm Ra}^{1/2} \approx h_0 {\rm Pr}^{3/2} {\rm Re}^2
\label{step1}
\end{equation}
On the other hand, Eqs.~(\ref{exactchi}) and (\ref{chiapprox}) imply
\begin{equation}
{\rm Nu} \approx {\rm Re} \  {\rm Pr}
\label{step2}
\end{equation}
Solving Eqs.~(\ref{step1}) and (\ref{step2}), we get
\begin{eqnarray}
{\rm Nu} &\approx& \frac{1}{h_0} {\rm Ra}^{1/2} {\rm Pr}^{1/2} \label{Nutheory}\\
{\rm Re} &\approx& \frac{1}{h_0} {\rm Ra}^{1/2} {\rm Pr}^{-1/2} \label{Retheory}
\end{eqnarray}
which are exactly the ultimate-state scaling observed~[see Eqs.~(\ref{Nuresult}) and (\ref{Reresult})].
Substituting Eqs.~(\ref{h}), (\ref{h0}), and (\ref{Retheory}) into Eq.~(\ref{dragapprox}), one gets
\begin{equation}
\epsilon_{drag} \approx \nu^3 k_0^4 h_0^2 {\rm Re}^3
\label{dragapproxF}
\end{equation}
which is in good agreement with our observation~[see Eq.~(\ref{epsilondrag})].

We have demonstrated that the presence of a drag that acts on the largest scales
is crucial for the observation of
the ultimate-state scaling of Nu and Re.
In our calculations and derivation, we employ a linear drag of the form $f u_0$.
An immediate question that arises is whether the scaling laws of Nu and Re 
depend on the specific mathematical form of the large-scale drag used. To answer this question,
we replace  $f u_0 \delta_{n,0}$ in Eq.~(\ref{undamp}) by
the general form of a nonlinear drag: $f u_0^{m-1} \delta_{n,0}$, where $m \ge 2$ is an integer, and
repeat our analysis to get a new estimate of the generalized average energy dissipation rate due
to the drag, ${\tilde \epsilon}_{drag}$. Now
\begin{equation}
{\tilde \epsilon}_{drag} = f \langle u_0^m \rangle \approx f \langle u_0^2 \rangle^{m/2} 
\label{generaldrag}
\end{equation}
Using Eq.~(\ref{Reapprox}), we get
\begin{equation}
{\tilde \epsilon}_{drag}
\approx \nu^3 k_0^4 h_0 
({\rm Ra}{\rm Pr}^{-1})^{\frac{3-m}{2}} {\rm Re}^m
\label{gdragapprox}
\end{equation}
where
\begin{equation}
f = h_0 (\sqrt{\alpha g \beta})^{3-m} k_0^{m-2}
\end{equation}
for some fixed value $h_0$ as Ra and Pr are varied.
Using $\epsilon_{total} \approx {\tilde \epsilon}_{drag}$, Eqs.~(\ref{exactepsilon}) and
(\ref{gdragapprox}), we now have
\begin{equation}
{\rm Nu} \ {\rm Ra}^{(m-1)/2} \approx h_0 {\rm Pr}^{(m+1)/2} {\rm Re}^m
\end{equation}
in place of Eq.~(\ref{step1}). Together with Eq.~(\ref{step2}),
which remains the same, we get
\begin{eqnarray}
{\rm Nu} &\approx& {h_0}^{-1/(m-1)} {\rm Ra}^{1/2} {\rm Pr}^{1/2} \label{NuGtheory}\\
{\rm Re} &\approx& {h_0}^{-1/(m-1)} {\rm Ra}^{1/2} {\rm Pr}^{-1/2} \label{ReGtheory}
\end{eqnarray}
In other words,
the scaling results of Nu~$\sim$~(RaPr)$^{1/2}$ and Re~$\sim$~(Ra/Pr)$^{1/2}$ remain valid for 
the general large-scale drag of $fu_0^{m-1}\delta_{n,0}$, for $m\ge 2$.

\section{Conclusions}
\label{summary}

An interesting question in turbulent thermal convection is how Nu and Re depend on Ra and Pr 
at very high Ra. Both the theories by Kraichnan~\cite{Kraichnan} and Grossmann and Lohse~\cite{GL} 
predicted that in this limit, Nu and Re would scale with Ra$^{1/2}$ for fluids with moderate Pr.
This kind of scaling behavior is taken to be the signature of the ultimate state of turbulent
convection.  However, the two theories have rather different assumptions about the role of the
boundary layers in heat transport. According to Kraichnan, 
convective eddies produced in the bulk generate
turbulent shear boundary layers and the turbulence of which enhances heat transport.
Thus we interpreted that in this picture of Kraichnan, 
buoyancy is directly acting on all scales in the bulk and that the boundary layers
play an important role in heat transport.
On the other hand, Grossmann and Lohse argued that in the limit of high Ra, 
the boundary layers would either break down or not contribute to the energy and 
dissipation and thus play no role in heat transport.
Studying numerically three-dimensional homogeneous turbulent convection 
in which buoyancy is acting only on the largest scales, 
Lohse and coauthors~\cite{LTPRL2003,bulkPF2005} 
reported scaling behavior of Nu and Re that is consistent with the ultimate-state scaling.

In the present work, we
have studied the scaling behavior of Nu and Re  
using a shell model of homogeneous turbulent convection in which buoyancy acts 
on most of the scales. In this model, buoyancy modifies the statistics of the velocity
fluctuations such that the statistics are given by Bolgiano-Obukhov plus corrections 
instead of Kolmogorov 1941 plus corrections~\cite{ChingCheng}. 
Moreover, there is an average 
inverse energy transfer from small to large scales such that
a large-scale drag has to be present for the system to achieve statistical stationarity.
Such a large-scale drag cannot exist by itself in the bulk of turbulent convection 
but could be resulted from the interaction of the inverse energy flow with
the boundaries. Thus, when buoyancy is acting directly on most of the scales in the bulk 
of turbulent convection, the boundary layers would play a crucial role and, as a result,
the flow cannot be bulk-dominated.
In this case, we have found that the dependence of Nu and Re on Ra and Pr 
is again consistent with the ultimate-state scaling. 

Because of the relative simplicity of the shell model, 
we can understand analytically the scaling behavior of Nu and Re.
The two exact results~Eqs.~(\ref{exactepsilon}) and (\ref{exactchi}) derived 
for the shell model are in exact analogy to those derived for 
turbulent Rayleigh-B{\'e}nard convection. As is clearly demonstrated by
the GL theory, to get the Nu and Re scaling, the whole task 
is to estimate $\epsilon_{total}$ and $\chi$.  In the shell model, 
there is always a cascade of entropy so that $\chi$ is given by $\kappa \beta^2$~RePr~[see
Eq.~(\ref{chiapprox})].
When buoyancy is acting on most of the scales, $\epsilon_{total}$ is 
dominated by $\epsilon_{drag}$. For a linear drag, $\epsilon_{drag}$ is
estimated as $\nu^3 k_0^4$ (Ra/Pr)$^{1/2}$Re$^2$~[see Eq.~(\ref{dragapproxF})]. 
We note that in this case, $\epsilon$ is given by 
$\nu^3 k_0^4 {\rm Re}^{5/2}$~[Eq.~(\ref{epsilontheory}) with $\eta\approx 3/5$ when
buoyancy acts on most of the scales] 
rather than the prediction of $\nu^3 k_0^4 {\rm Re}^3$ by the GL theory.
Putting these estimates into the two exact results, we get the ultimate-state scaling of
Nu~$\sim$~(RaPr)$^{1/2}$ and Re~$\sim$~(Ra/Pr)$^{1/2}$. On the other hand, 
when buoyancy is acting only as a driving force on the largest scales, 
$\epsilon_{drag}$ is negligible and $\epsilon_{total}$ given by $\epsilon$ as usual. 
In this case, the statistics of the temperature resemble those of a passive scalar,
and $\epsilon$ is given by $\nu^3 k_0^4 {\rm Re}^3$~[Eq.~(\ref{epsilontheory}) with
$\eta\approx 1/3$ when buoyancy acts only on the largest scales], the usual result 
obtained in inertia-driven turbulence without buoyancy, which is also the result 
derived in the GL theory. In this case Eqs.~(\ref{GLepsilonShell}) and (\ref{GLchiShell}) 
hold, leading to the ultimate-state scaling as shown in the GL theory.

Hence there are two different physical scenarios that can give rise to 
the ultimate-state scaling of Nu~$\sim$~(RaPr)$^{1/2}$ and Re~$\sim$~(Ra/Pr)$^{1/2}$ 
for fluids with moderate Pr. In the first scenario, which is illustrated in the present work, 
buoyancy is acting directly on most of the scales of the bulk of turbulent convection and,
on average, energy is transferred from small to large scales. 
An effective damping at the largest scales, which can be provided by
the interaction of the inverse energy transfer with the boundaries, is crucial. 
In the second scenario, buoyancy is acting 
only as a driving force on the largest scales, temperature in the bulk of the convective 
flow is behaving like a passive scalar statistically, and the boundary layers play 
no role in heat transport. The first scenario is in accord with the physical picture
presented in Kraichnan's work~\cite{Kraichnan} 
while the second scenario is in accord with that proposed by the GL theory~\cite{GL}.
The next question would be whether or not buoyancy acts directly 
on most of the scales in the bulk of turbulent Rayleigh-B{\'e}nard convection 
at very high Ra, and the answer of which would help to distinguish which scenario
is physically relevant.

\acknowledgments
The authors thank Roberto Benzi for stimulating discussions and K.R. Sreenivasan
for reminding us of Ref.~\cite{DoeringConstantin}.
This work is supported in part by the Hong Kong Research Grants
Council (Grant No. CA05/06.SC01).

\end{document}